# Trusted Data Forever: Is AI the Answer?

Emanuele Frontoni[1], Marina Paolanti[1], Tracey P. Lauriault[2], Michael Stiber[3], Luciana Duranti[4] and Muhammad Abdul-Mageed[5]

[1] *VRAI Vision Robotics and Artificial Intelligence Lab, University of Macerata, Italy*
[2] *Critical Media and Big Data Lab, Carleton University, Ottawa, ON K1S 5B6, Canada*
[3] *Intelligent Networks Lab, University of Washington Bothell, WA, USA*
[4] *InterPARES Lab, University of British Columbia, Vancouver, BC V6T 1Z4, Canada*
[5] *NLP and ML Lab, University of British Columbia, Vancouver, BC V6T 1Z4, Canada*

**Abstract**

Archival institutions and programs worldwide work to ensure that the records of governments, organizations, communities, and individuals are preserved for future generations as cultural heritage, as sources of rights, and as vehicles for holding the past accountable and to inform the future. This commitment is guaranteed through the adoption of strategic and technical measures for the long-term preservation of digital assets in any medium and form — textual, visual, or aural. Public and private archives are the largest providers of data big and small in the world and collectively host yottabytes of trusted data, to be preserved forever. Several aspects of retention and preservation, arrangement and description, management and administrations, and access and use are still open to improvement. In particular, recent advances in Artificial Intelligence (AI) open the discussion as to whether AI can support the ongoing availability and accessibility of trustworthy public records. This paper presents preliminary results of the InterPARES Trust AI ("I Trust AI") international research partnership, which aims to (1) identify and develop specific AI technologies to address critical records and archives challenges; (2) determine the benefits and risks of employing AI technologies on records and archives; (3) ensure that archival concepts and principles inform the development of responsible AI; and (4) validate outcomes through a conglomerate of case studies and demonstrations.

## 1. Introduction

Archival institutions and programs worldwide work to ensure that the records of governments, organizations, communities, and individuals are preserved for future generations as cultural heritage, as sources of rights, to hold the past accountable, and as evidence to inform future plans. A record – or archival document – is any document (i.e. information affixed to a medium, with stable content and fixed form) made or received in the course of an activity, and kept for further action or reference. Because of the circumstances of its creation a record is a natural by-product of activity, is related to all the other records that participate in the same activity, are impartial with respect to the questions that future researchers will ask of them, and are authentic as instruments of activity. This is why records are inherently trustworthy. Thus, their preservation must ensure that any activity carried out on the records to identify, select, organize, describe them and make them accessible to the people at large must ensure that they remain trustworthy, that is reliable (i.e. their content can be trusted), accurate (i.e. the data in them are unchanged and unchangeable), and authentic (i.e. their identity and integrity are intact). This is particularly difficult in the digital environment because the content, structure and form of records are no longer inextricably linked as they used to be in the traditional records environment [1][1]. The issue exists for both digital and digitized records and is rendered more serious by the sheer number and volume of records that have accumulated overtime and are being created today in a large variety of systems.

The InterPARES (International research on Permanent Authentic Records in Electronic System) project has addressed these issues since 1990, focusing on current and emerging technologies as they evolve and developing theory, methods, and frameworks that allow for the ongoing preservation of the records resulting from the use of such technologies[2]. The latest iteration of InterPARES, *I Trust AI*, funded, as previous projects, by the Social Sciences and Humanities Research Council of Canada, but differs as it is not concerned with the records produced by a specific technology, but has the purpose of using AI to carry out archival functions for the control in the long term of all records, on any medium, and from any age, and to do so in such a way that the trustworthiness of the records remains protected and verifiable, and that the tools and processes are transparent, unbiased, equitable, inclusive, responsible (i.e. protecting autonomy and privacy) and sustainable[3]. There have been several projects looking at AI in archives, but they typically look at a particular tool in a specific context or even a single set of records, and they tend to use off the shelf tools. The research question that *I Trust AI* is being asked here is: "what would AI look like if archival concepts, principles and methods were to inform the development of AI tools?" What is lacking is comprehensive, systematic research into the use of AI to carry out the different archival functions in an integrated way and ensure the continuing availability of verifiable trustworthy records to prevent the erosion of accountability, evidence, history and cultural heritage. Thus, we are addressing the technological issues from the perspective of archival theory, by integrating the technology with complex human-oriented tools. The objectives of *I Trust AI* are to 1. Identify specific AI technologies that can address critical records and archives challenges; 2. Determine the benefits and risks of using AI technologies on records and archives; 3. Ensure that archival concepts and principles inform the development of responsible AI; and 4. Validate outcomes from Objective 3 through case studies and demonstrations. Our approach is two-pronged, comprising the practical and immediate need to address large-scale existing problems, and the longer-term need to have AI-based tools that are reliably applicable to future problems. Our short-term approach focuses on identifying high impact problems and limitations in records and archives functions, and applying AI to improve the situation. This will be achieved via collaboration between records and archival scientists and professionals and AI researchers and industry experts. Our long-term approach focuses on identifying the tools that records and archives specialists will need in the future to flexibly address their ever-changing needs. This includes decision support and, once decisions are made, rapid implementation of AI-based solutions to those needs.

---

[1] "The Concept of Record in Interactive, Experiential and Dynamic Environments: The View of InterPARES". Archival Science 6, 26-33.
[2] www.InterPARES.org

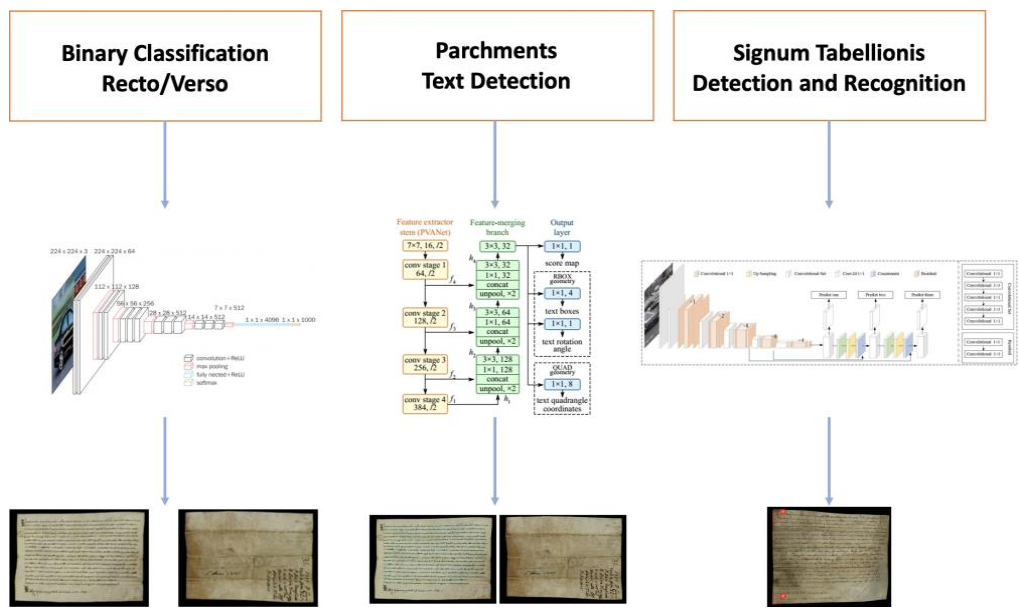

**Figure 1:** PergaNet DL pipeline. PergaNet DL pipeline consists of three stages: classification of parchments recto/verso, the detection of text, then the detection and recognition of the "signum tabellionis". Firstly, a VGG16 Network trained on a dataset of scanned parchments is needed to solve a classification task: recto/verso. After, the text in the image is detected. Then, YOLOv3 was used to predict bounding box locations and classify these locations in one pass.

The *I Trust AI* project is a multinational interdisciplinary endeavour, and this means that our first effort must be to understand each other, starting with the language we use. For example, archival professionals talk about records, while computer scientists and AI professionals talk about data. To the former data are the small-est meaningful unit of information in a record. To an AI specialist, data are arrangements of information (possibly in a database), be these facts or not, regardless of their size, nature and form.. Thus, for the purposes of this paper, which is directed to data analytics specialists, we will use the term data.

---

[3] www.interparestrustai.org

**Table 1**
Digitalised Heritage Data

| Digitalised Heritage Data | Size |
|---|---|
| Fondo Ufficio italiano brevetti e marchi, Trademarks series: volumes with trademark registrations | 30 TB |
| Official collection of laws and decrees | 15 TB |
| Fund A5G (First World War): files with various documents (reports, reports, correspondence) | 1 TB |
| Special collections (documents declassified under the Renzi and Prodi Directives): reports, reports, circulars | 2 TB |
| Judgments of military courts | 3 TB |
| Various photographic funds | 2 TB |
| Digitised study room inventories | 15 TB |
| National Archives of the US | 1323 TB |

Public and private archives are the largest providers of data big and small in the world as they collectively host yottabytes of trusted data, to be preserved forever. Their creators are organizations and individuals from myriad sectors and disciplines, from public administration, to academia and businesses of all kind (e.g. banking, engineering, architecture, gaming), and from Indigenous communities, civil society organizations, associations, and virtual communities. Table 1 reports the data quantity. The Italian State Central Archives, for example, has Data quantity (67 TB) of "digital objects" stored by the ACS linked to the typology (digitised heritage data (in TIFF)). The National Archives of the US has 1,323 terabytes of electronic data. This paper presents some preliminary results of I Trust AI international research partnership and is organized as follows: Section 2 provides a general discussion of AI and its subsets; Section 3 describes three of the about forty studies that are now in course; and Section 4 presents an overview of the kind of studies that are being pursued at this time and a conclusion.

## 2. Artificial Intelligence and Deep Learning

There are various definitions of what AI is. For example, Russel and Norvig [2] define AI as a field focused on the study of intelligent *agents* that perceive their *environment* and take *actions* to maximize their chance of *success* at some *goal*. In recent times, however, AI has been used much more widely to refer to any technology where there is some level of automation especially resulting from the application of deep learning in various domains. Deep learning (DL) is a sub-field of machine learning (ML), which are both sub-fields of AI. Similar to AI, ML is defined in several ways. A classical definition of ML comes from Mitchell and Learning [3] who provide a procedural definition maintaining that a computer program is said to learn from *experience E* with respect to some class of *tasks T* and *performance measure P* if its performance at tasks in *T*, as measured by *P*, improves with experience *E*. As to DL [4, 5], it is a class of ML methods inspired by information processing in the human brain. What makes DL powerful is its ability to automatically learn useful *representations* from data. DL algorithms, unlike classical ML methods, are not only able to learn the mapping from representation to output but also to learn the representation itself [6], thus alleviating the need for costly human expertise in crafting features for models. DL has achieved success in recent years in a wider variety of applications in many domains involving various types of data modalities such as language, speech, image, and video.

As mentioned, DL mimics information processing in the brain. This is possible by designing artificial neural networks arranged in multiple layers (and hence the term *deep*) that take input, attempt to learn a good representation of it, and map it to an output decision (e.g., *is this text in Greek or Latin?*). The way these networks are designed can vary, thus, various types of deep learning architectures have been proposed. Two main types of DL architectures have been quite successful. These are *recurrent neural networks (RNNs)* (e.g., [7]), a family of networks specializing in processing sequential data, and *convolutional neural networks (CNNs)* [8], an architecture specializing in data with a 'grid-like' typology (e.g., image data) [5]. More recent advances, however, abstract away from these two main types toward more dynamic networks such as the Transformer [9].

In general, ML and DL methods learn best when given large amounts of labeled data (e.g., for a model that detects sensitive information, labels can be from the set *sensitive, not-sensitive*). DL in particular is data-hungry and tends to learn best given large amounts of labeled data. This type of learning with labeled data is called *supervised learning*. It is also possible to work with smaller labeled datasets. In these cases, training samples can be grown iteratively exploiting unlabeled data based on decisions from an initial model (*self-training*) or using decisions from various initial models (*co-training*). This is called *semi-supervised learning* [10, 11, 12]. The third main-type of ML methods is *unsupervised learning* where a model usually tries to cluster the data without access to any labels. There are other paradigms such as *distant supervision* where a model attempts to learn from surrogate cues in the data in absence of high-quality labels (e.g., [13]). *Self-supervised learning* where real-world data are turned into labeled data by masking certain regions (e.g., removing some words or parts of an image) and tasking a model to predict the identity of these masked regions is currently a very successful approach. These various methods of supervision can also be combined to solve downstream tasks. For example, Zhang and Abdul-Mageed [14] combine self-supervised language models with classical self-training methods to solve text classification problems. The next section will introduce three *I Trust AI* studies.

## 3. Case Studies in I Trust AI

### 3.1. Data from Emergency Services Communications Systems

One of the cornerstones of of public safety and societal wellbeing is a reliable and comprehensive emergency services communications system (ESCS, such as 9-1-1 in the US and Canada). Such systems can be considered to encompass the organizations, electronic infrastructure, and policies and procedures that enable answering and responding to emergency phone calls [15]. As might be expected of systems that originate in analog, switched telephony, ESCS evolution into a digital system has resulted in a haphazard conglomeration of subsystems and generally needs re-imagining as a modern technological solution. In the US, this change has been termed the "Next Generation 911" (NG911) project.

A transformation such as NG911 once again re-casts ESCS as keystone information and communication technology, subject to all of the concerns of such systems: cybersecurity, privacy, crisis preparedness, strategic and operational decisions, etc. At the same time, it opens up possibilities for data analytics to improve ESCS performance, inform funding decisions, monitor the health of societies and their infrastructures, and serve as early warnings for natural and human made crises. This study connects large- scale simulations of ESCS to historical data from ESCS operations to develop and document an understanding of how to preserve authentic, reliable data that can be used for applications such as re-creation of past events (as might be done to support training or to explore the effects of changes in policies and procedures), testing system operation in one locale based on data from another, and examining how various data analytics techniques might be usefully applied — understanding what characteristics of these data might be usable to researchers Specifically, we are addressing the following research questions:

**What real-world and simulation ESCS data are available to be preserved for access by researchers?** The answer to this varies greatly from locale to locale, depending on technology in use, public policy, and controlling agency procedures. Moreover, for such data to be available, we must understand privacy and security risks associated with transferring them from their current owners to a research environment, along with the risk of misinterpreting them if they are decontextualized from whatever tacit knowledge might exist within the owning organizations. We are also considering pragmatic issues, such as building a knowledge base of legal restrictions on collection in various jurisdictions, formal processes for collecting these data (such as data sharing agreements), variation in the culture of practice surrounding such data, potential biases that might result from systematic differences in different areas' capacity to collect and share data (such as might arise from regional funding differences), and understanding the metadata and other information (such as ESCS physical and operational structure and, generally, the policies and practices that determine what and how data are generated and collected).

**What are the challenges and benefits of discovering knowledge patterns from historical ESCS data?** These patterns will serve as clues to developing protocols for ESCS managers to follow regarding data collection, and as clues for how these data can be applied for reuse. We will consult with external stakeholders to seek advice and to run thought experiments using surveys, think-aloud exercises, retrospective first-hand accounts, etc. We will also examine historical records of disasters for which the preserved data are more complete than basic ESCS datasets.

**What other data/metadata associated with emergency events are not part of the ESCS data stream?** From our preliminary examination, typical ESCS data currently involve lists of individual calls and information directly associated with such calls (perhaps including full or partial phone numbers, call categorization, GPS coordinates, responder information, response times, etc.). What these datasets do not directly include are events and data that are external to the call stream but are the reason for such calls (traffic, weather, geopolitical events, and so on). Some of these additional data may be present in other sources of information in a format that can be reasonably collected in tandem with call data. On the other hand, it may be the case that other causal information must be inferred from available data (and, of course, it may be the case that a combination of inference and extraction from other data streams could be useful). For the inferencing task, we propose using an examination of simulation results and simulation artifact provenance information as exemplars to develop a set of specifications for what an AI-driven system would need to accomplish [16].

**What are the roles of the disciplines of Archival Science and Artificial Intelligence in building a central repository for ESCS data?** Both the individual fields of archival science and artificial intelligence, plus their overlap or combination that could be considered to fall within the realm of data science, have a number of roles in the organization and interpretation of ESCS data. We provide examples below from the application of real-world data to simulations:

- Generating requirements for simulator design so that simulation output matches real data in terms of format, metadata, etc.
- Analyzing and comparing simulation output with real- world data.
- Synthesizing ESCS data that match features of real-world data as part of an overall ESCS simulation.
- Using real-world call data to drive a simulation of an emergency response system, for example, to allow a "replay" of a previous disaster or to investigate how modifications to such a system might produce different outcomes.

### 3.1.1. Progress

This work-in-progress is in its initial stages, preparing for the point in time when we can begin collecting ESCS data. Specifically, we are:

- working with a small set of external partners to develop a general understanding of ESCS operations, policies, and procedures, as well as identifying which data exist;
- developing a process within our project for working with a selected ESCS management organization to build an understanding of their specific operations and data;
- fleshing out a model data sharing agreement to serve as a starting point for discussions surrounding transferring data to our research environment;
- consulting with our institutional review board regarding the use of this particular set of human data;
- and configuring a secure internal data storage system.

Subsequently, we will prepare an initial case study in which we will apply the above to a single locale: we will go through all of the steps of understanding the ESCS processes that produced the data, developing a data sharing agreement, and collecting data and metadata.

### 3.2. Learning from Parchments

The digitization of historical parchments is extraordinarily convenient, as it allows easy access to the documents from remote locations and removes the need for the possible adverse effects of their physical management and access [17]. This arrangement is particularly suitable to archives and museums who preserve such invaluable historical documents whose contents are unpublished and which, if damaged, cannot be fully restored by conventional tools, are difficult to read on the original, due to high levels of damage and the delicate nature of the material. Damaged parchments are notably prevalent in archives all over the world [18]. Their digital representations reduce both damage and access issues by providing users with the possibility of reading their contents at any moment, from remote locations, and without necessitating the potentially harmful physical handling of the document. Thus, the automatic analysis of digitized parchments has become an important research topic in the fields of image and pattern recognition. It has also been a considerable research issue for several years, gaining attention recently because of the value that maybe be unlocked by extracting the information stored in historical documents [19]. Interest in applying AI/ML to ancient image data analysis is becoming widespread, and scientists are increasingly using this method as a powerful and complex process for statistical inference. Computer-based image analysis provides an objective method of identifying visual content independently of subjective personal interpretation, while potentially being more sensitive, consistent and accurate than physical human analysis. Learned representations often result in much better performance than hand-designed representations when it comes to these types of texts. Until now, parchment analysis has required physical user interaction, which is very time consuming. Hence, the effective automatic feature extraction competence of Deep Neural Networks (DNNs) decreases the demand for a personal physical extraction processes.

Considering the above, PergaNet is a lightweight DL-based system for the historical reconstructions of ancient parchments is specifically designed and developed for this type of analysis. The aim of PergaNet is to automate the analysis and processing of large volumes of scanned parchments. This problem has not yet been deeply investigated by the computer vision community as parchment scanning technology is still novel, but it has proven to be extremely effective for data recovery from historical documents whose content is inaccessible due to the deterioration of the medium. The proposed approach aims to reduce hand-operated analysis while using manual annotations as a form of continuous learning. The whole system however requires digital labour, such as the manual tagging of large training data. Up until now, large datasets remain necessary to boost the performance of DL models, and manually verified data will be used as continuous learning and maintained as training datasets. PergaNet comprises three important phases: the classification of parchments recto/verso, the detection of text, and the detection and recognition of the "sig,num tabellionis". (i.e. the identifier of the author). PergaNet concerns not only the recognition and classification of the objects present in the images, but also their location. This *I Trust AI* study expands the implementation of AI guided by archival institutions and programs as this method could be used by many other archives for different types of documents. The analysis is based on data about the ordinary use by researchers of this type of material and does not involve altering or manipulating techniques aimed to generate data. This provides actionable insights that are helpful to identify text as documentary form and not as reading.

The DL pipeline is depicted in Figure 1. We chose VGG16 Network [20] for its suitability and effectiveness in image classification tasks and were inspired by the work of Zhou et al. [21], in the way in which PergaNet detects the text in the image. This phase allows for the exclusion of the text on the parchment in the phase of recognition of the signa. The DNN model chosen is EAST for word detection [21]. Finally, a Convolutional Neural Network has been employed for the signa detection. Our approach uses YOLOv3 [22], an algorithm that processes images in real time. We chose this algorithm because of its efficiency in computational terms and for its precision to detect and classify objects. The network is pre-trained using COCO[4], a publicly available dataset; this was a choice made to reduce the need for a large amount of training data, that would come with a high computational cost.

### 3.3. Digital Twin Study

Spatial media [23] and spatial data infrastructures (SDI) [24] have normalized as complex interconnected global, regional, national, and personalized social and technological systems of systems. Simply consider the monitoring of climate at a global scale to inform the logistics of production chains, or to predict, preempt and prevent disaster resulting from natural calamities on local physical infrastructure or simply to report the temperature and humidity levels outside to your smart phone to help you plan your day. This occurs seamlessly in the background of our daily activities and involves a vast complex of sensors; databases; cloud computing centres; telecommunication networks and the internet; standards; code, software and platforms; people and institutions; laws, regulation and policies; communication systems and of course AI/ML [25].

An emerging subset of these spatial data infrastructures are digital twins (DTs). A DT is an ecosystem of multi-dimensional and interoperable subsystems made up of physical things in the real-world, digital versions of those real things, synchronized data connections between them and the people, organizations and institutions involved in creating, managing, and using these[5]. In terms of physical and real things consider a building or a car manufacturing plant; a digital representation of those things in a digital platform or an interactive virtual reality game engine; with an internet of things (IoT) system of sensors and databases that communicates between the buildings or manufacturing processes in the plant in real and near real time and the people and institutions that own and operate these. Contemporary examples of DTs are the modeling and managing of the construction of Sweden's new high speed rail systems[6]; Hyundai car and ship manufacturing plants[7]; and as part of smart city strategies (see the submissions to the Infrastructure Canada Smart City Challenge[8]). DTs originate in the aerospace industry, first with NASA's Apollo 13 in the 1970s, although in that case it was a physical replica to help troubleshoot issues of a ship in flight; and were predominantly used in manufacturing and logistics [26] Increasingly, DTs involve building information modeling (BIM) such as REvit a proprietary platform or the OS BlenderBIM; whereby a building will be conceptualized and rendered into a 3-dimensional drawing with attributes captured in a database; often replacing typical blue prints. The BIM informs the construction of the building; and is a record of it once completed. BIM renderings are increasingly being submitted as part of the building permit approval process (BuildingSmart, 2020, e-submission common guidelines for introducing BIM to the building process[9]) and are updated into as is BIMs for ongoing operations. BIMs are also used to estimate mate-rial costs as they are interconnected with material vendor databases and electrical and heating systems. BIMs interrelate with smart asset management systems (ASM) which inform building maintenance and operations, and monitor internal climate such as temperature, humidity, air flow and quality; which are inputs for the AI/ML systems that remotely manage heating and cooling systems; electricity use and consumption as well as maintenance schedules and inform ongoing decision making.

---

[4]https://cocodataset.org/#home

[5] CIMS, 2021,About page, what is a digital twin https://canadasdigitaltwin.ca/about-2-2/

[6]Pimental, K. 2019, Visualizing Sweden's first high-speed railway with real-time technology, https://www.unrealengine.com/en-US/spotlights/visualizing-sweden-s-first-high-speed-railway-with-real-time-technology

[7]Chang-Won, L., 2022, Hyundai Motor works with Unity to build digital-twin of factory supported by metaverse platform; https://www.ajudaily.com/view/20220107083928529

[8] https://www.infrastructure.gc.ca/sc-vi/map-applications.php

[9] https://www.buildingsmart.org/wp-content/uploads/2020/08/e-submission-guidelines-Published-Technical-Report-RR-2020-1015-TR-1.pdf

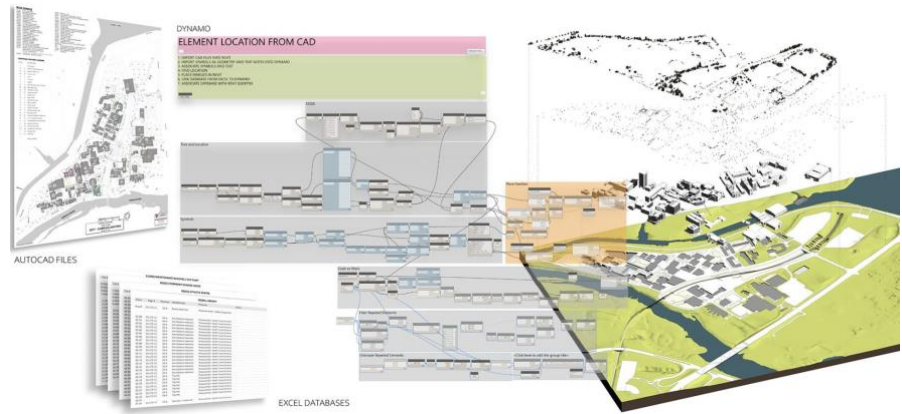

**Figure 2:** Integrating diverse databases into BIM (Image created by Nico Arellano, CIMS 2021).

In this *I Trust AI* Digital Twin case study we collaborate with researchers working on the Imagining the Canada Digital Twin (ICDT)[10] project funded by the Canadian New Frontiers in Research Fund[11] which proposes a national, inclusive, and multidisciplinary research consortium for the creation of a technical, cultural, and ethical framework to build and govern the technology, data and institutional arrangements of Canada's DT. ICDT focuses on the built environment, concentrating on the Architecture, Engineering, Construction, and Owner Operator (AECOO) industry. ICDT is led by the Carleton Immersive Media Studio (CIMS)[12] developing a DT prototype of the Montréal-Ottawa-Toronto corridor using a simulated, distributed server network. The research study involves an interdisciplinary research team of architects, data scientists, engineers, building scientists, archival professionals and critical data studies scholars, from Carleton University CA, Luleå University of Technology SE, the Swedish Transport Administration and the University of Florence IT who will develop a Use and Creation preservation case study. The Study aims to preserve the DT of campus buildings and structures created as part of SUSTAIN (https://cims.carleton.ca/#/projects/Sustain) and the Carleton Digital Campus Innovation (DCI)[13] project integrating Building Performance Simulation (BPS) technologies with BIM on a campus scale; building information management systems (BIM), Asset Management Systems (AMS), visualizations of the digital structures in the Unreal Game Engine, VR and modelling, AI/ML, and Real-time data for decision making.

The Carleton Campus DT data of seven buildings belong to the University that must preserve these as official records, which are used for operations by Facilities, Management and Planning (FMP) and are part of research and development for the CIMS and SUSTAIN projects; thus involving research data that must be managed and deposited in a trusted digital repository.

The implications of this research are important to the archival community who will increasingly have to ingest complex record sets such as these, as well as create an archival package to maintain the integrity of these complex interlinked DT systems through time. This study will be one of the first globally to examine the preservation of a DT. Its research questions are: Can a digital twin be preserved and what is required at the point of creation to ensure that it can be? Can information about the AI tools, automation and real time data involved in this complex data, social and technological system be preserved, and how? And, what might be the role of AI/ML be in terms of creating an archival package to ingest a digital twin? The outputs of this research will provide empirical data to meet the objectives of the I Trust AI Project; and also provide Carleton University with the opportunity to test the preservation of Campus DT records in its institutional archives. In the process, it will inform the technology sectors involved in the creation of DTs, so that they may build-in at the point of creation, the necessary bread crumbs for long term preservation.

---
[10] https://canadasdigitaltwin.ca/
[11] https://www.sshrc-crsh.gc.ca/funding-financement/nfrf-fnfr/index-eng.aspx
[12] https://cims.carleton.ca/#/home
[13] https://www.cims.carleton.ca/#/projects/DigitalCampusInnovation

## 4. Conclusions and Future Works

The studies presented above are only 3 of about 40 in-progress studies which cover a wide range of subjects and issues, such as enterprise master data management, the preservation of AI techniques as paradata, modelling an AI-assisted digitization project, gamification of archival experience for users, declassification of personal information using AI tools, and user approaches and behaviours in accessing records and archives in the perspective of AI. The challenges we are addressing with this project have never before been systematically and globally dealt with; it is enormous and fraught, but critical. While the risks of using AI to solve the problems of managing the ever-growing, ever-more-diverse bodies of public and private records throughout their lifecycle, from creation to preservation and access, are unknown, the risks of not acting in concert to do so are unacceptable: loss of the ability to secure people's rights, of evidence of past acts and facts to serve as a foundation for decision making, and of historical memory.

This project will significantly impact society in several areas. (1) Records-keeping in local and national government agencies is a vital part of our society's ability to maintain oversight on and accountability of governance, but, with the inability to handle the vast quantities of digital records, public bodies risk undermining their own legitimacy as oversight if they can not appropriately process and make accessible information in a timely fashion. By helping address this crisis through the development, evaluation, and contextualization of AI techniques we contribute to the ability of agencies and institutions to maintain their place in our democracies. (2) Automation techniques can potentially aid in the economic viability of many cash-starved records offices and archival institutions by ensuring that professional records management and archival expertise are used wisely, with classification tools and TAR able to allow a quick review and assessment of vast quantities of records. Similarly, with businesses depending on records agencies for routine activities, improved speed in responding to queries will bring a positive effect to the economy. (3) AI techniques have the potential to aid in the accessibility of records in archives by new audiences, for instance by translating and indexing historical materials written in indigenous languages, sensitising problematic archival descriptions, helping patrons find connected items, or captioning historical photographs. These techniques have both a cultural significance, by providing better access to historical material, and a social and scientific significance, by making current records easier to organise, retrieve and use by both their creators and the public at large. (4) While there have been numerous calls to action to systematically explore the application of AI techniques to the records and archives field, AI also currently faces major ethical challenges that will benefit from an archival theory perspective, for instance in dealing with bias and personal information. By exploring further the connections between AI and archives, this project is and will contribute to the intellectual progress of both fields. The *I Trust AI* project has generated a great amount of enthusiasm among participant researchers (about 200) and partner organizations (87), as well as organizations that do not have the capacity to participate but look forward to outcomes they can use, because it deals with issues that are already dramatically changing the way we act, behave and think. We have a unique and essential contribution to make, because we have the means of creating knowledge ensuring that digital data and records are controlled and made accessible in a trustworthy, authentic form wherever they are located; are promptly available when needed; duly destroyed when required; and accessed only by those who have a right to do so.